\documentclass[aps,prc,twocolumn,floatfix]{revtex4}
\usepackage{epsfig}
\usepackage{amsmath}
\def\be{\begin{equation}}
\def\ee{\end{equation}}

\begin{document}

\title{Projection and ground state correlations made simple}
\author{K. Hagino$^1$, P.-G. Reinhard$^{2,3}$, G.F. Bertsch$^2$}
\affiliation{$^1$: Yukawa Institute for Theoretical Physics, Kyoto
University, Kyoto 606-8502, Japan}
\affiliation{$^2$: Institute for Nuclear Theory and Department of
Physics, University of Washington, Seattle, Washington 98195, USA}
\affiliation{$^3$: Institut f\"ur Theoretische Physik II,
         Universit\"at Erlangen-N\"urnberg,\\
         Staudtstrasse 7, D--91058 Erlangen, Germany}

\date{\today}

\begin{abstract}
We develop and test efficient approximations to estimate ground state
correlations associated with low- and zero-energy modes.  The scheme
is an extension of the generator-coordinate-method (GCM) within
Gaussian overlap approximation (GOA). We show that GOA fails in
non-Cartesian topologies and present a topologically correct
generalization of GOA (topGOA). An RPA-like correction is derived as
the small amplitude limit of topGOA, called topRPA.  Using exactly
solvable models, the topGOA and topRPA schemes are compared with
conventional approaches (GCM-GOA, RPA, Lipkin-Nogami projection) for
rotational-vibrational motion and for particle number projection.
The results shows that the new schemes perform very well in all
regimes of coupling. 
\end{abstract}

\maketitle

\section{Introduction}

Self-consistent mean-field models are nowadays the standard tool for
nuclear structure calculations. Their quality has reached a level
where one needs to take into account correlation effects beyond mean
field, particularly those which are related to low-energy- or
symmetry-modes. Typical examples are center-of-mass projection,
particle-number projection, angular-momentum projection or quadrupole
surface vibrations. There is a large variety of techniques to deal
with those correlations, for a review see \cite{correlrev}.  The most
widely used ones are the random phase approximation (RPA), see
e.g. \cite{Rowe,BB94}, and the generator coordinate method (GCM), see
e.g. \cite{GCM,heenenGCM}. The latter has close links to projection
formulae. The RPA has the advantage that it provides simple equations
because it employs only second order commutators of the basic one-body
operators with the Hamiltonian. However, it runs into difficulties
with soft modes which arise typically near transition points. The GCM
is very general and extremely robust, but also very cumbersome to
handle. Thus one has developed simplifications in the aim to use also
preferably second order expressions. This is achieved by the Gaussian
overlap approximation (GOA) to GCM, for details see the review
\cite{GCM}. GCM-GOA is a fair compromise between the generality of GCM
and the simplicity of RPA. It uses up to second order anti-commutators
but can still deal with large amplitude collective motion. Second
order approximation within the spirit of GOA have also been widely
applied to projection schemes. The standard recipe $E_{\rm
cm}=\langle\hat{P}_{\rm cm}^2\rangle/2mA$ for center of mass
correction belongs to this class \cite{schmid}.  The similarly simple
rotational correction $\langle\hat{J}^2\rangle/2\Theta$ has been
widely employed, e.g. in the large scale fits of \cite{goriely}.  And
there is the well known Lipkin-Nogami approach for
particle-number projection \cite{Pra73a}.

However, one has to be aware that the GOA is not always performing
well. For example, it fails for rotational motion in weakly deformed
systems and for particle number projection in the regime of weak
pairing. The failure can be related to the topology of the collective
coordinate under consideration. GOA is well suited for Cartesian
coordinates which extend in the interval $(-\infty,+\infty)$ with
constant volume element. The best example is here center of mass
motion. But GOA is not necessarily appropriate for other topologies
as, e.g., rotational motion whose coordinates are defined on a
sphere. It can still work if the overlaps are falling off very
quickly. But regimes of weak coupling have broad overlaps and thus the
topology of the underlying coordinates is fully explored. It needs to
be built in into the approximation. An example for rotational motion
is found in \cite{rot}. A most general construction for any topology
is discussed in \cite{pom}. The changes are, in fact, obvious and
simple. It amounts to building the topology of the coordinates into the
parameterization of the GOA. We call the emerging approach a
topologically corrected Gaussian overlap approximation (topGOA).

It is the aim of this paper to investigate the accuracy of the topGOA
for two cases most relevant in nuclear structure calculations,
deformations and particle-number projection.  We compare
topGOA with RPA as well as full GCM and simple GOA.  Furthermore, we
derive a small amplitude limit of topGOA which gives at the end
very simple and compact formulae for the collective ground-state
correlations, in a sense comparable to RPA. We call that approach
topRPA. In both test cases we employ a suitable generalization of the
Lipkin-Meshkov-Glick model \cite{LMG}.

\section{Conventional approaches}

This section provides a brief summary of traditionally well known
approaches for collective correlations, RPA and GCM up to GOA.

\subsection{RPA correlations}
\label{sec:RPAcorr}

The RPA theory is perhaps the most straightforward treatment of correlations
beyond mean field theory.  It gives the leading corrections in the limit
of large number of interacting particles.  With the RPA, one calculates
an excitation spectrum of eigenfrequencies $\omega_n$ and the associated
particle-hole operators  $\hat{C}^+_n$ that generate the eigenmodes.
These modes are also present in the RPA ground state as zero-point motion,
leading to a RPA theory of the ground state correlation energy, see e.g. 
\cite{Rowe,ring}.  For a single mode, the
RPA
correlation energy is given by

\begin{equation}
  \Delta E
  =
  \frac{\omega}{2}
   \left(1 - \langle\Phi_0|\{\hat{C},\hat{C}^+\}|\Phi_0\rangle\right)
\label{eq:RPAcorr}
\end{equation}

where
\begin{equation}
  \omega
  =
  \langle\Phi_0|[\hat{C},[\hat{H},\hat{C}^+]]|\Phi_0\rangle
\end{equation}
and  $|\Phi_0\rangle$ is the mean field  ground
state. In the case the mode corresponds to a broken continuous symmetry,
$\omega=0$ and the formula should be applied by taking the 
$\omega\longrightarrow 0$ limit.
It is also advantageous in that case to
separate the generators into time-even and time-odd 
\begin{equation}
\label{eq:PQdef}
  \hat{Q}
  =
  \frac{1}{\sqrt{2}}\left(\hat{C}^++\hat{C}\right)
  \quad,\quad
  \hat{P}
  =
  \frac{i}{\sqrt{2}}\left(\hat{C}^+-\hat{C}\right)
  \quad.
\end{equation}
The $\hat{P}$ is usually the generator of a collective deformation,
for example a center-of-mass shift in case of the translational mode. 
Particle-number
projection is an example where the time-even operator $\hat{N}$ spans
the collective space.

The RPA correlation energy (\ref{eq:RPAcorr}) can fail due to
double counting if one employs a sum over a large RPA spectrum
\cite{fukuda64}, but double counting is negligible if only a few
collective modes are used \cite{rowe68}. That is the line of approach
followed here. For a most recent survey of RPA correlations along that
line, see \cite{hagber}. It will be taken up explicitly in the
applications later on.

\subsection{Generator coordinate method}

\subsubsection{General framework}

The most general technique for constructing collective modes is 
the generator coordinate method (GCM). It utilizes a superposition
of wave functions defined along some
collective deformation path
$\left\{|\Phi_q\rangle\equiv|q\rangle\right\}$. Each state $|q\rangle$
along this path is an independent particle state (or independent
quasi-particle state in case of BCS). The correlated wave function 
is given by
\begin{equation}
  |\Psi\rangle
  =
  \int dq\,|q\rangle\,f(q)
  \quad.
\end{equation}
where the superposition function $f$ is determined by the
Griffin-Hill-Wheeler  equation
\begin{subequations}
\label{eq:GHW}
\begin{eqnarray}
  &&
  \int dq'\,\left[{\cal H}(q,q')-E{\cal I}(q,q')\right]f(q')
  =
  0
  \quad,
\\
 && 
 {\cal H}(q,q')
  =
  \langle q|\hat{H}|{q'}\rangle
  \quad,
\\
  &&
  {\cal I}(q,q')
  =
  \langle q|{q'}\rangle
  \quad .
\end{eqnarray}
\end{subequations}
Normalizing $\Psi$, the
correlation energy is given by
\begin{equation}
  \Delta E
  =
  \langle\Psi|\hat{H}|\Psi\rangle
  -
  \langle\Phi_0|\hat{H}|\Phi_{0}\rangle
\end{equation}
where $|\Phi_{0}\rangle\equiv|0\rangle$ is the ground state of the
underlying independent particle model.
The GCM
can be easily generalized to multiple modes. One simply generalizes
$q$ to a vector of deformations and extends $\int dq'$ to a
multi-dimensional integral, see e.g. \cite{JanSch,brink68,GCM}.  The
GCM is often applied in this straightforward, but tedious, manner
where the overlaps and the solution of the Griffin-Hill-Wheeler equation are determined
numerically, see e.g. \cite{Egi89a,heenenGCM,heenenGCM2}.

\subsubsection{The Gaussian Overlap Approximation (GOA)}

The full GCM is much more elaborate than RPA because one deals with
the overlaps for any combination of $q$ and $q'$ and the highly
non-local Griffin-Hill-Wheeler equation. A dramatic simplification is achieved by the
Gaussian overlap approximation (GOA). It represents the dependence of the
overlaps on the difference $(q-q')$ by a Gaussian times a
polynomial in $(q-q')^n$.  The overlap is represented as a pure
Gaussian,
\begin{equation}
  {\cal I}(q,q')
  =
  \exp{\left(-\frac{\lambda}{4}(q-q')^2\right)}  
  \quad,
\end{equation}

with

\begin{equation}
  \lambda(\bar{q})
  =
  \frac{1}{2}\left(i\partial_q-i\partial_{q'}\right)^2{\cal I}(q,q')
\Big|_{q=q'=\bar{q}}
  \quad,
\end{equation}
One usually goes up to second order
derivatives in the expression for the Hamiltonian
\begin{subequations}
\begin{eqnarray}
\label{eq:GOA}
  \frac{{\cal H}(q,q')}{{\cal I}(q,q')}
  &=&
  {\cal H}_0(\bar{q})
  -
  \frac{1}{8}(q-q')^2{\cal H}_2(\bar{q})
  \quad,
\\
  {\cal H}_0(\bar{q})
  &=&
  {\cal H}(\bar{q},\bar{q})
  \quad,
\\
  {\cal H}_2(\bar{q})
  &=&
  \left(i\partial_q-i\partial_{q'}\right)^2
  \left.\frac{{\cal H}(q,q')}{{\cal I}(q,q')}\right|_{q=q'=\bar{q}}
  \quad,
\\
  \bar{q}
  &=&
  \frac{q+q'}{2}
  \quad.
\end{eqnarray}
\end{subequations}
For further details, see \cite{GCM}.  The GOA yields a dramatic
simplification of the Griffin-Hill-Wheeler equation. Assuming that the
coefficients depend only weakly on $\bar{q}$, one can recast the
Griffin-Hill-Wheeler
equation into a collective Schr\"odinger equation with a simple second
derivative as operator for the kinetic energy. Large amplitudes in
average collective deformation $\bar{q}=(q+q')/2$ are still
allowed. Thus the GCM-GOA is applicable to conditions of large
fluctuation as are typical for low-energy modes and for symmetry
projection.

A further dramatic simplification emerges if one restricts the
considerations to small amplitudes also in $\bar{q}=(q+q')/2$. 
Then collective dynamics becomes harmonic and all expressions can
be worked out analytically. The final result is then just the RPA
\cite{JanSch,rowe68,brink68}. The correlations from GCM-GOA become then
identical to the RPA correlations as given the above subsection
\ref{sec:RPAcorr}.

\subsubsection{Beyond GOA}

However, the GOA has its limitations. The Gaussian ansatz assumes
tacitly that the collective coordinate spans the interval
\begin{equation}
  q\in(-\infty,+\infty)
  \quad.
\end{equation}
In other words, the dynamics is fundamentally Cartesian in the collective
coordinates. This is certainly true for some situations, e.g. the center-of-mass motion
where each coordinate $R_x$, $R_y$, and $R_z$ runs over
$(-\infty,+\infty)$. But the presupposition is violated in many
cases. In particular, in rotational motion the rotation angles are
restricted to finite intervals with periodic boundary conditions.
For such situations the GOA can be generalized by modifying the
the arguments of the Gaussian to correctly
include the topology of the collective mode \cite{rot,pom}. We call this
generalization the
topological GOA (topGOA). The details of topGOA depend, of course, on
the actual mode under considerations. In the following, we exemplify
and test topGOA for two typical and most important applications in
nuclear physics: deformations and particle-number
projection. The projection is straightforward and yields immediately
expressions in second order throughout. The efficient treatment of
deformations remains an important problem in nuclear structure.  The
theory should provide accurate correlation energies, going from the 
small-amplitude vibrational limit to the large
amplitude static deformations and including the soft region in between.
These applications will serve as
a critical testing ground for the topRPA, and the small amplitude
approximation to topGOA.

\section{Vibrations and rotational projection}

\subsection{The three-level model}

The usual two-level Lipkin-Meshkov-Glick Hamiltonian 
has been widely used to model
the collective motion of in a deformation coordinate, as it contains 
the vibrational and static deformation limits with the mean-field phase
transition in between.  However, the model does not have a continuous
symmetry, which is an important aspect of the deformations. To include
a continuous symmetry, we have extended the space in the 
Lipkin-Meskov-Glick model to
three levels, and call the extended model the three-level model. Two of the
levels are degenerate in the three-level, and the interactions treats those
levels identically.  This 
introduces a symmetry mode with the topology of rotations in a plane.
For clarity we repeat here the definition of the three-level model; for details 
see \cite{hagber}.
The three levels are labelled 0,1, and 2. The basic $1ph$ transitions
$0\!\rightarrow\!1$ are induced by $\hat{K}_{+,1}$ and those to state
2 by $\hat{K}_{+,2}$. The amount of excitation is measured by
$\hat{K}_{0,i}\,,\,i\in\{1,2\}$. The $\hat{K}$ operators obey a
quasi-spin algebra.  The Hamiltonian of the model reads 
\begin{subequations}
\begin{eqnarray}
  \hat{H}
  &=&
  \epsilon\sum_{i=1}^2\hat{K}_{0,i}
  -
  \chi\frac{\epsilon}{2(N-1)}\sum_i(\hat{K}_{+,i}^2+\hat{K}_{-,i}^2)
  \quad,
\label{eq:hamLMG3}\\
  \hat{K}_{0,i}
  &=&
  \sum_{m=1}^N\alpha_{im}^+\alpha_{im}^{\mbox{}}
  \quad,
\\
  \hat{K}_{+,i}
  &=&
  \sum_{m=1}^Na_{im}^+a_{0m}^{\mbox{}}
  \quad,
\\
  \hat{K}_{-,i}
  &=&
  \sum_{m=1}^Na_{0m}^+a_{im}^{\mbox{}}
  \quad.
\end{eqnarray}
\end{subequations}
The exact solution of this Hamiltonian is obtained by diagonalization
in the space of $\hat{K}_x^n\hat{K}_y^m$. 
The three-level Hamiltonian
is the first term in eq. (\ref{eq:hamLMG3}). It defines the energetic
relations amongst the levels. Note that the two excited states $i=1,2$
are degenerate. This gives the model the rotational symmetry.
The second term in (\ref{eq:hamLMG3}) models a two-body interaction.
It is again symmetric in $i=1\leftrightarrow 2$ which maintains
rotational symmetry. The strength is regulated by $\chi$, defined 
to be dimensionless coupling strength. We will see later that
$\chi\sim 1$ is the critical point in the model separating weak and
string coupling. 

It is convenient to analyze the many-particle wave function in terms
of collective variables $\alpha$ and $\beta$.  The collective wave
function is defined
\begin{equation}
  |\alpha\beta\rangle
  =
  e^{\tan(\alpha)\hat{K}_+(\beta)}|0\rangle{\cal N}^{-1/2}(\alpha)
  \quad,
\label{eq:collbasis}
\end{equation}
where
\begin{equation}
  \hat{K}_+(\beta)
  =
  \cos(\beta)\hat{K}_{+,1}+\sin(\beta)\hat{K}_{+,2}
  \quad,
\end{equation}
and the normalization is given by
\begin{equation}
  {\cal N}(\alpha)
  =
  \langle 0|e^{\tan(\alpha)\hat{K}_-}e^{\tan(\alpha)\hat{K}_+}|0\rangle
  =
  \cos^{-2N}(\alpha)
  \quad.
\end{equation}
Note that the model is rotationally invariant in the angle
$\beta$. The motion in $\alpha$ corresponds to collective
vibrations. The system is close to a good vibrator for small residual
interaction, $\chi<1$. It is a rigid rotator for large $\chi>1$.  The
transitional regime $\chi\sim 1$ explores collective motion with large
amplitude fluctuations.
Two subtle details need to be mentioned: First, there is only one
rotational degree-of-freedom which means that the model corresponds to
rotations in a plane.  Second, the vibrational degree-of-freedom
contains relevant information only in the interval
$\alpha\in\{0,\pi\}$, similar to the vibrational mode in the usual
Lipkin-Meshkov-Glick model. This is the price one pays to have a
simple model.

The simplicity of the model allows one to write down the exact overlaps
analytically:
\begin{widetext}
\begin{subequations}
\begin{eqnarray}
  {\cal I}(\alpha\beta,\alpha'\beta')
  &=&
  {\left[\cos(\alpha)\cos(\alpha')
         +\sin(\alpha)\sin(\alpha')\cos(\beta\!-\!\beta')\right]^N}
  \quad,
\\
  \frac{{\cal H}(\alpha\beta,\alpha'\beta')}
       {{\cal I}(\alpha\beta,\alpha'\beta')}
  &=&
  N\epsilon\,
  \frac{\sin\alpha\sin\alpha'\cos(\beta-\beta')}
  {\cos\alpha\cos\alpha'+\sin\alpha\sin\alpha'\cos(\beta-\beta')} 
  -\chi\frac{\epsilon}{2}N\,
  \frac{\sin^2\alpha\cos^2\alpha'+\cos^2\alpha\sin^2\alpha'}
  {(\cos\alpha\cos\alpha'+\sin\alpha\sin\alpha'\cos(\beta-\beta'))^2}
  \quad.
\end{eqnarray}
\label{eq:exovLMG3}
\end{subequations}
\end{widetext}

The Hartree-Fock (HF) solution is obtained simply by minimizing the
expectation value of the Hamiltonian in a state $|\alpha\beta\rangle$,
\begin{eqnarray}
  E_{\rm mf}(\alpha)
  &=& {\cal H}(\alpha\beta,\alpha\beta) \nonumber \\
  &=&
  N\epsilon\sin^2(\alpha)
  -
  \chi\epsilon N\sin^2(\alpha)\cos^2(\alpha)
  \quad.
\end{eqnarray}
with respect to the deformation $\alpha$,$\beta$. This yields the
Hartree-Fock energy as $E_{\rm HF}=E_{\rm mf}(\alpha_{\rm HF})$ where the
deformation of the minimum is denoted by $\alpha_{\rm HF}$.  Note
that the energy is independent of the actual value of $\beta$ due to
rotational invariance of the three-level model.

\subsection{The RPA modes}

Small amplitude motion around the HF minimum induces collective
excitations of the system. They can be worked out analytically for the
three-level model \cite{hagber}. There are two collective modes to be
considered.  At spherical shape $\alpha_{\rm HF}\sim 0$, there are two
degenerate vibrational modes. The degeneracy is lifted with increasing
$\alpha_{\rm HF}$.  With further increasing $\alpha_{\rm HF}$, there
comes a critical point where the RPA solutions become unstable.  A
different scenario develops after the transition point.  The two modes
separate into a rotational mode along $\beta$ and a vibrational mode
along $\alpha$.  The two eigenfrequencies are $\omega=0$, associated
with the rotational mode, and $\omega=\epsilon\sqrt{\chi^2-1}$ for
vibrations.  Having these two modes at hand, one can compute the RPA
correlation energy applying eq. (\ref{eq:RPAcorr}) for each mode
separately and add up the result to the total correlations.

\subsection{The topGOA for the three-level model}

The standard GOA overlaps can be obtained by expanding
eq. (\ref{eq:exovLMG3}) with respect to $(\alpha-\alpha')$ and 
$(\beta-\beta')$ up to the second order. 
We exemplify it here for the norm kernel at $\alpha=\alpha'$ 
and expansion in $\beta-\beta'$. GOA reads 
\begin{eqnarray*}
{\cal I}(\alpha\beta,\alpha\beta')
&=&
  {\left[\cos^2(\alpha)
         +\sin^2(\alpha)\cos(\beta\!-\!\beta')\right]^N}, \\
&=&\left[1-2\sin^2(\alpha)\sin^2\left(\frac{\beta-\beta'}{2}\right)
\right]^N, \\
&\longrightarrow&\exp\left(-\frac{N}{2}\sin^2(\alpha)(\beta-\beta')^2\right). 
\end{eqnarray*}
The problem is obvious: the exact overlap is periodic in $\beta-\beta'$
while the GOA is not. 

To develop an appropriate ansatz for topGOA we have to look at the
topology of the collective coordinates.
The pair of coordinates $(\alpha,\beta)$ extends over the surface of
the unit sphere. The exact overlaps (\ref{eq:exovLMG3}) hint already
the combinations of coordinates which is generated by this topology:
$\cos(\alpha)\cos(\alpha')
         +\sin(\alpha)\sin(\alpha')\cos(\beta\!-\!\beta')$.
It is the measure for a distance on the sphere.
The idea of topGOA is to apply to the norm overlap the Gaussian limit
theorem for the shape of the overlap function while preserving the
topological combination of the arguments. Similar combinations are to
be used for expanding the Hamiltonian overlap. This yields then for
the three-level model the form
\begin{subequations}
\label{eq:topGOArot}
\begin{eqnarray}
  {\cal I}(\alpha\beta,\alpha'\beta')
  &=&
  e^{-\frac{\lambda_\alpha}{4}
            \sin^2\left(\frac{\alpha\!-\!\alpha'}{2}\right)
            -\frac{\lambda_\beta}{4}
         {\cal S}_\alpha\sin^2\left(\frac{\beta\!-\!\beta'}{2}\right)}
  \;,
\label{eq:basicGOA}\\
  \lambda_\alpha
  &=&
  \frac{1}{2}\left(i\partial_\alpha-i\partial_{\alpha'}\right)^2
   {\cal I}(\alpha\beta,\alpha'\beta)\Big|_{\alpha=\alpha'=\bar{\alpha}}
\\
  \lambda_\beta{\cal S}_\alpha
  &=&
  \frac{1}{2}\left(i\partial_\beta-i\partial_{\beta'}\right)^2
   {\cal I}(\alpha\beta,\alpha\beta')\Big|_{\beta=\beta'=\bar{\beta}}
\\
  \frac{{\cal H}(\alpha\beta,\alpha'\beta')}
       {{\cal I}(\alpha\beta,\alpha'\beta')}
  &=&
   {\cal H}_0
   -
   {\cal H}_2^\alpha\sin^2\left(\frac{\alpha\!-\!\alpha'}{2}\right)
\nonumber\\
  &&
   -
   {\cal H}_2^{\beta}{\cal S}_\alpha
  \sin^2\left(\frac{\beta\!-\!\beta'}{2}\right)
  \quad,
\label{eq:hamover}\\
  {\cal H}_0
  &=&
  {\cal H}(\bar{\alpha}\bar{\beta},\bar{\alpha}\bar{\beta})
\label{eq:collpot}\\
  {\cal H}_2^\alpha
  &=&
  \frac{-1}{2}
  \left(\partial_\alpha\!-\!\partial_{\alpha'}\right)^2
  \left.
  \frac{{\cal H}(\alpha\beta,\alpha'\beta)}
       {{\cal I}(\alpha\beta,\alpha'\beta)}
  \right|_{\alpha=\alpha'=\bar{\alpha}} 
\end{eqnarray}
\begin{eqnarray}
  {\cal H}_2^\beta
  &=&
  \frac{-1}{2{\cal S}_\alpha}
  \left.
  \left(\partial_\beta\!-\!\partial_{\beta'}\right)^2
  \frac{{\cal H}(\alpha\beta,\alpha\beta')}
       {{\cal I}(\alpha\beta,\alpha\beta')}
  \right|_{\beta=\beta'=\bar{\beta}}
\\
  {\cal S}_\alpha
  &=&
  \bar{s}^2
  -
  \sin^2\left(\frac{\alpha\!-\!\alpha'}{2}\right)
\nonumber\\
  \bar{s}
  &=&
  \sin(\bar{\alpha})
  \quad,\quad
  \bar{c}
  =
  \cos(\bar{\alpha})
  \quad,
\nonumber\\
  \bar{\alpha}
  &=&
  \frac{\alpha+\alpha'}{2}
  \quad.
\nonumber
\end{eqnarray}
\end{subequations}
Thus far we have the topGOA overlaps for any system where the
collective coordinates form the topology of a sphere. The specific
coefficients for the present three-level model are
\begin{subequations}
\begin{eqnarray}
  \lambda_\alpha
  &=&
  8N
\\
  \lambda_\beta
  &=&
  8N
\\
  {\cal H}_0
  &=&
  N\epsilon\bar{s}^2
  -
  N\chi\epsilon \bar{s}^2\bar{c}^2
  \quad,
\\
  {\cal H}_2^\alpha
  &=&
  N\epsilon(\bar{c}^2-\bar{s}^2)+N\chi\epsilon (4\bar{s}^2\bar{c}^2+1)
  \quad,
\\
  {\cal H}_2^\beta
  &=&
  2N\epsilon\bar{c}^2+4N\chi\epsilon \bar{s}^2\bar{c}^2
  \quad.
\end{eqnarray}
\end{subequations}

\begin{figure}[bht]
\centerline{\epsfig{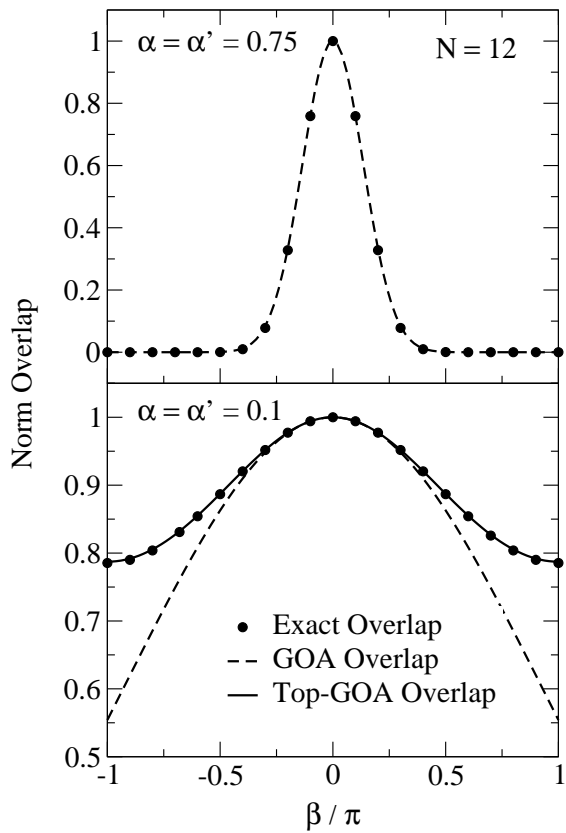}}
\caption{\label{fig:normov}\sl 
The norm overlaps along $\beta$ direction for two different
deformations $\alpha$ as indicated,
for the three-level model with $N=12$. The dots show the exact overlaps,
dashed lines stand for the standard GOA and full line for topGOA.
The topGOA is not shown in the upper panel because it graphically
identical with standard GOA.
}
\end{figure}

The effect of GOA versus topGOA for the norm overlap is demonstrated
in figure \ref{fig:normov}.  For large deformations (upper panel), the
norm overlap decays rather quickly in angle $\beta$.  The conventional
GOA is here a reliable approximation. The situation is much different
at small deformation.  The overlaps become broad and hit the
periodicity limits.  This yields a dramatic difference between GOA and
topGOA. Note that topGOA is still an excellent approximation to the
exact overlap while GOA fails badly.

\subsection{Performance of topGOA}

The conventional GOA (\ref{eq:GOA}) maps the Griffin-Hill-Wheeler equation
(\ref{eq:GHW}) into a collective Schr\"odinger equation of second
order order in the collective momentum \cite{GCM,heenenGCM}.  This
feature is lost in topGOA. Further approximation steps would be needed
to come to that end.  We will not pursue them further here and solve
directly the Griffin-Hill-Wheeler equation (\ref{eq:GHW}) inserting
the topGOA overlaps (\ref{eq:topGOArot}).

\begin{figure}[t!]
\centerline{\epsfig{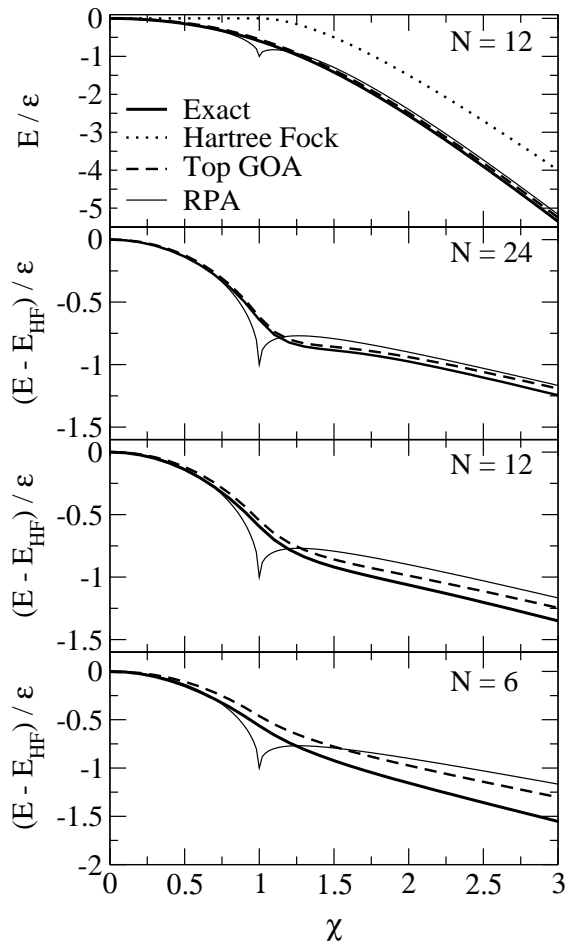}}
\caption{\label{fig:LMG3_comp}\sl 
Comparison of energies for the three-level model
at various level of approximations as indicated
Upper block: Total energies for $N$=12.
Lower block:
The correlation energy $\Delta E =
E-E_{\rm HF}$ for various N as indicated.
Results are drawn for a large range of coupling strengths $\chi$ from
sphericity $\chi=0$ deeply into the deformed regime.
}
\end{figure}
Figure \ref{fig:LMG3_comp} compares the RPA and topGOA
with HF and the exact result for a large variety of coupling
strengths. The uppermost panel shows total energies. One sees that
both approaches correct the HF energy very far towards the exact
energy. However the RPA shows irregularities near the critical point
$\chi\approx 1$.

A more detailed look is given in the three lower panels of figure
\ref{fig:LMG3_comp} where we show the correlation energies $\Delta
E=E-E_{\rm HF}$ for the various approaches and for a series of system
sizes. The RPA provides a useful correction in the limits of
sphericity and well developed deformations, but fails badly around the
critical point. The topGOA performs very well in all regimes. The
results improve with increasing system size as one could expect from
the Gaussian limit theorem inherent in topGOA. Acceptable results are
obtained from topGOA also for $N$=4. But all approaches become inaccurate
for
$N$=2 which is obviously not collective enough.

\subsection{Angular momentum projection}

When the mean-field ground state breaks a symmetry of the Hamiltonian,
one can get an improved wave function and energy by projection, i.e.
take a minimal set of states $q$ and appropriate $f$ in eq. (4) to
enforce the symmetry.  This is particularly useful for deformations
and projection of the $J=0$ ground state ground state out of a
deformed intrinsic state.  The questions before us are, how
does this technique compare with the RPA or the topGOA for 
computing the correlation energy?  It should be noted that the
projection method has a formal advantage in that the calculated energy
is an upper bound of the true energy associated with the Hamiltonian.

\subsubsection{The projected state}

We will examine how well the projection technique works for 
the three-level model as test case. Rotational projection on the
ground state angular momentum $M=0$ reads simply
\begin{equation}
  |\alpha\rangle_{\rm proj}
  \propto
  \int_{-\pi}^{\pi}d\beta\,|\alpha\beta\rangle
  \quad.
\end{equation}
The rotationally projected energy is computed as expectation value
which amounts to integrating the overlaps over the angular coordinate
$\beta$, i.e.
\begin{equation}
  E_{\rm proj}(\alpha)
  =
  \frac{\int d(\beta-\beta'){\cal H}(\alpha\beta,\alpha\beta')}
       {\int d(\beta-\beta'){\cal I}(\alpha\beta,\alpha\beta')}
  \quad.
\label{eq:projE}
\end{equation}
This is simple and straightforward for the topGOA overlaps of the form
(\ref{eq:topGOArot}). We thus can skip the details.

\subsubsection{Variation before and after projection}
\label{sec:VAP}

The energy (\ref{eq:projE}) can be computed for any given deformation
$\alpha$.  The HF ground state deformation $\alpha_{\rm HF}$ is
obtained from minimizing the mere HF energy (\ref{eq:collpot}).
Applying the projection on this state corresponds to the scheme
``variation-before-projection''. It serves to correct for the angular
momentum fluctuations in the deformed HF ground state. A much better
approach is obtained when performing ``variation-after-projection''
\cite{ring}. Here one minimizes the projected energy (\ref{eq:projE}).
This is an involved task for exact projection. The topGOA approach
yields a simple expression for the projected energy on which a
variation is still feasible.  It is, of course, particularly simple in
the present test case.  We just have to search for the deformation
$\alpha_{\rm proj}$ which minimizes $E_{\rm proj}$.

\begin{figure}[t!]
\centerline{\epsfig{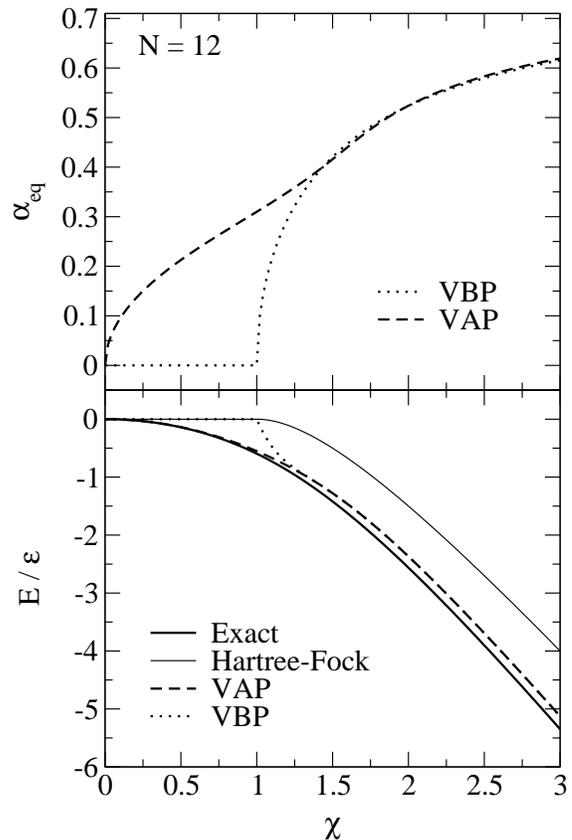}}
\caption{\label{fig:demo_VAP}\sl 
Comparison of variation-after-projection (VAP) and
variation-before-projection (VBP) in the three-level model with $N$=12
using exact projection.  Upper panel: Ground state deformation
$\alpha_{\rm eq}$.  Lower panel: Ground state energies from HF and
from rotational projection both ways, compared with the exact energy.
}
\end{figure}
Variation-before-projection and variation-after-projection are
compared in figure \ref{fig:demo_VAP} for a large range of coupling
strengths. The upper panel shows the ground state deformations. The
variation-before-projection state stays spherical up the critical
point and switches to deformation with a discontinuous derivate
(second order transition). The variation-after-projection states
develop more smoothly and show a steady growth of deformation.
variation-after-projection can afford intrinsic deformations because
it ``knows'' that projection will restore spherical symmetry.  The
freedom which variation-after-projection exploits will yield a lower
energy. This is shown in the lower panel of figure \ref{fig:demo_VAP}.
It is obvious that variation-after-projection picks up a large
fraction of the correlation energy at any coupling strength $\chi$,
80\% for strongly deformed systems and even more for weakly deformed
ones.  This makes it obvious that variation-after-projection is the
superior strategy. Mind that topGOA helps to simplify
variation-after-projection considerably.  We will test it now in the
next paragraph.

\subsubsection{Performance of topGOA for a.m. projection}

\begin{figure}[t!]
\centerline{\epsfig{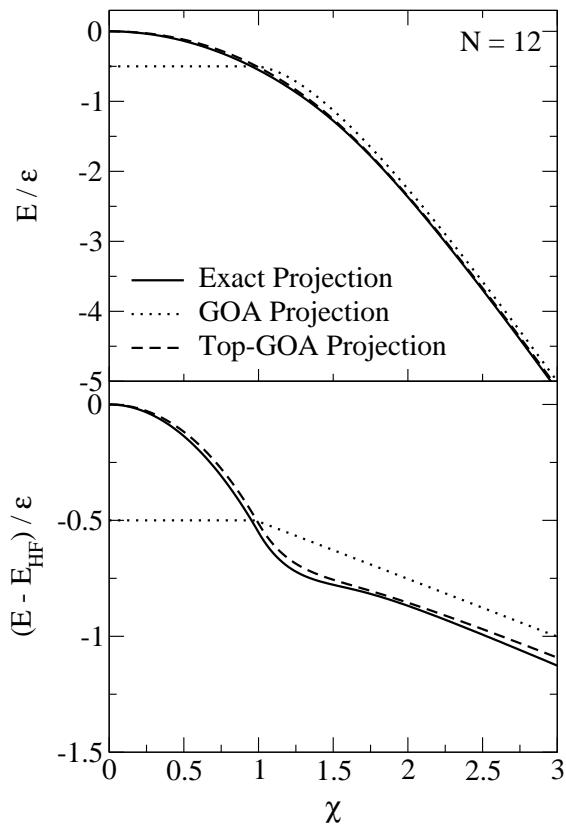}}
\caption{\label{fig:comp_prjrot}\sl 
The correlation energy $\Delta E =
E_{\rm HF}$ at various levels of approximation for the three-level 
model N=12. The compared cases are:
full = exact projection energy, dotted = projection using standard GOA,
dashed = projected energy using topGOA. The upper panel shows the
energies
as such and the lower panel shows the correlation energy, i.e. the
difference to the mere HF energy.
}
\end{figure}
The performance of topGOA for rotational projection is checked in
figure \ref{fig:comp_prjrot} for the case of $N$=12.
Conventional GOA has obviously problems at small deformation up to
beyond the critical point. But topGOA provides a very good
approximation to exact projection throughout. And it does that on the
grounds of a simple expression for the projected energy which can be
deduced from second order moments only. This is welcome for an
efficient variation-after-projection and it is particularly helpful in connection with
effective energy functionals because double (anti-)commutators with
$\hat{H}$ can still be safely derived from second functional
derivatives, see Sec. IV E.

The rotational projection can still be done with second order
information around the minimum point. It is thus as simple to compute
as RPA. And this simple part provides the dominant portion of the
correlation energy.  The most costly part of the correlation energy is
computing
the small final contribution from vibrations. It is tempting to
consider mere angular projection as a first guess for the correlation
energy. That is, in fact, a strategy pursued in the large scale fits
of \cite{goriely}. Our result here provides a welcome substantiation of
their ``rule of thumb''.

\subsection{A thoroughly second order approach: topRPA}

The conclusions from the previous subsection encourage the quest for
a more efficient estimator
of the vibrational correlation energy. And the typical
pattern of variation-after-projection add reasons to that. We have seen in figure
\ref{fig:demo_VAP} that the variation-after-projection ground state is nearly always
deformed. The projected energy as function of $\alpha$ has always a
fairly well developed minimum much in contrast to the HF energy which
is rather soft around the critical point. This hints that one is
allowed to perform a small amplitude expansion about the projected
minimum $\alpha_{\rm proj}$. Once having accepted this idea,
the remaining steps are obvious and simple:
\begin{enumerate}
 \item 
  One performs variation-after-projection using topGOA for rotational projection.
  This yields the variation-after-projection ground state deformation $\alpha_0$.
 \item  
  One computes the topGOA projected energy 
  \begin{equation}
    E_{\rm proj}(\bar{\alpha})
    =
    E_{\rm proj}({\alpha}_0)
   +\frac{1}{2}(\bar{\alpha}-\alpha_0)^2\partial_{\bar{\alpha}}^2 
   E_{\rm proj}(\alpha_0)
  \end{equation}
  in the vicinity of $\alpha_0$ and deduces the curvature 
  $\partial_{\bar{\alpha}}^2 E_{\rm proj}$ 
  of this effective potential.
 \item
  For the remaining vibrational correction, one applies the 
  simple correlation energy from harmonic approximation
  \begin{subequations}
  \begin{eqnarray}
    \delta E_{\rm vib}
    &=&
    \frac{1}{2}\sqrt{\partial_{\bar{\alpha}}^2 E_{\rm proj}{\cal B}}
    -
    \left(
    \frac{\partial_{\bar{\alpha}}^2 E_{\rm proj}}{4\lambda_{\rm proj}}
    +
    \frac{\lambda_{\rm proj}{\cal B}}{4}
    \right)
    \quad,
  \\
    {\cal B}
    &=&
    \frac{2{\cal H}_2^\alpha(\alpha_0)}{\lambda_{\rm proj}^2}
    \quad.\\
\lambda_{\rm proj} &=&
2\partial_\alpha\partial_{\alpha'}\langle\alpha'|\alpha\rangle _{\rm proj}
\Big|_{\alpha=\alpha'=\alpha_0}, \\
{\cal H}^\alpha_2(\alpha_0)&=&
\left.2\partial_\alpha\partial_{\alpha'}
\frac{\langle\alpha'|\hat{H}|\alpha\rangle _{\rm proj}}
{\langle\alpha'|\alpha\rangle _{\rm proj}}
\right|_{\alpha=\alpha'=\alpha_0}.
  \end{eqnarray}
  \end{subequations}
 \item
   The total energy is then finally 
   $$
     E=\delta E_{\rm proj}(\alpha_0)- \delta E_{\rm vib}
     \quad.
   $$
\end{enumerate}
Note that this scheme requires only information on second order
derivatives in $\alpha$ and $\beta$ about the deformed ground state.
In that sense it is much similar to RPA. We thus call that scheme
topologically corrected RPA (topRPA). The essence is, of course, that
topological constraints are exploited to construct from the given
second order information the final ground state energy in topRPA.

\begin{figure}[t!]
\centerline{\epsfig{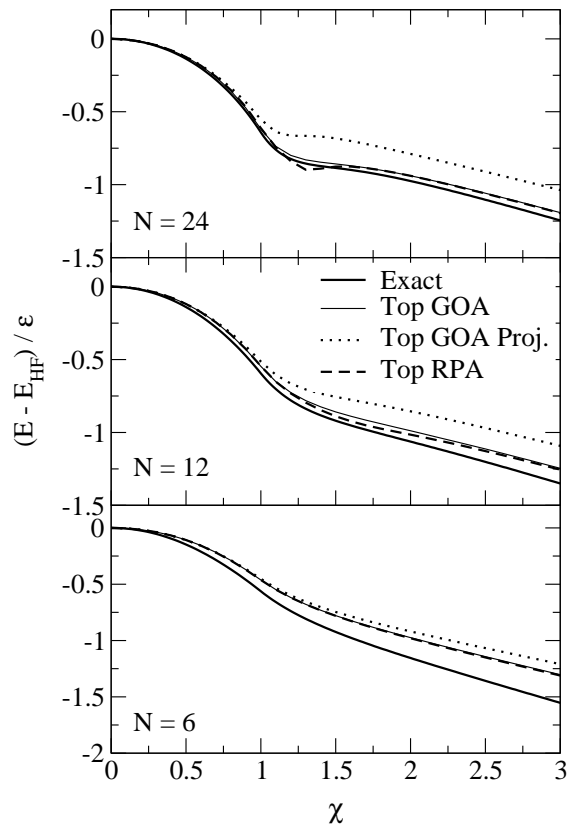}}
\caption{\label{fig:comp_topRPA}\sl
Comparison of topGOA and topRPA for the three-level model and various N as
indicated.
Upper block: total energies compared with HF and the exact result.
Lower block: 
The correlation energies $E_{\rm scheme}-E_{\rm HF}$ for various
levels of approximation. ``scheme'' stands for the exact solution
(full line), the topGOA (dotted line), the topRPA (dashed line). 
Results are drawn versus effective
coupling strength $\chi$. 
}
\end{figure}
Figure \ref{fig:comp_topRPA} compares the performance of topRPA and
topGOA for the correlation energy in the three-level model. It is obvious
that topRPA provides a good approximation to topGOA, equally good for
all system sizes. Both schemes constitute a reliable approach to the
exact result, better for larger systems. For completeness, we show
also the correlation from angular momentum projection alone. We see
again that this exhausts the leading part of correlations and could be
considered as quick and simple approach. However, topRPA is not much
more expensive and comes close to the final result.

\section{Particle-Number-Projection}
\label{sec:PNP}

The second test cases in this paper is concerned with particle-number
projection. It becomes necessary when starting Hartree-Fock-Bogoliubov
(HFB) states, or its BCS approximation, are involved. The HFB approximation
produces independent quasi-particle states which have mixed particle
number $N$. One needs to project the HFB states onto good particle
number. This is important in any nuclear structure calculation because
doubly magic nuclei (where mere HF suffices) are an extremely rare
species. Similar as in the previous example of rotations-vibrations
there is, in principle, a pair of modes, namely particle number
projection and pairing vibrations. We confine here the discussion to
projection alone because that is the widely used strategy and because
it will again exhaust the dominant part of the correlations.

\subsection{Exact projection}

Let $|\Phi_0\rangle$ be a HFB state with average particle number
\begin{equation}
  \langle\Phi_0|\hat{N}|\Phi_0\rangle
  =
  N_0
  \quad.
\end{equation}
The projected state with exact particle number $N_0$ is
\begin{subequations}
\begin{equation}
  |\Psi\rangle
  \propto
  \int_0^{2\pi}d\eta\,|\eta\rangle
  \quad,\quad
  |\eta\rangle
  =
  e^{i\eta\tilde{N}}|\Phi_0\rangle
\label{eq:PNPpath}
\end{equation}
where 
\begin{equation}
  \tilde{N}
  =
  \hat{N}-N_0
  \quad.
\end{equation}
\end{subequations}
The construction of the path from straightforward
$\exp{(i\eta\tilde{N})}$ makes the norm overlap a function of the
difference alone, i.e. ${\cal I}={\cal I}(\eta-\eta')$.  The number
conservation $[\hat{H},\tilde{N}]=0$ causes also ${\cal H}={\cal
H}(\eta-\eta')$. The projected energy thus becomes
\begin{equation}
  E
  =
  \frac{\int d\eta\langle\Phi_0|\hat{H}e^{i\eta\tilde{N}}|\Phi_0\rangle}
       {\int d\eta\langle\Phi_0|e^{i\eta\tilde{N}}|\Phi_0\rangle}
  =
  \frac{\int d\eta\,{\cal H}(\eta)}{\int d\eta\,{\cal I}(\eta)}
  \quad.
\label{eq:Eproj}
\end{equation}

\subsection{The topGOA for particle number projection}

\subsubsection{Overlaps and correlation energy}

The collective path is $|\eta\rangle$ as given in
eq. (\ref{eq:PNPpath}).  The collective coordinate is defined in the
interval $[0,2\pi)$ and is periodic as
$\eta\longrightarrow\eta+2\pi$. This periodicity is not reproduced by
the standard GOA overlaps (\ref{eq:GOA}).  One has to modify GOA to
account for that structure, in other words one has to employ
topologically correct GOA (topGOA).  Taking up the experience from the
previous test case, we can postulate that the periodic structure of
the coordinates is properly taken into account by 
the argument in GOA through
$$
  \frac{\eta}{2}
  \longrightarrow
  \sin\left(\frac{\eta}{2}\right)
  \quad.
$$
One may wonder why we use this particular assignment for the
for the
generalization.  The choice is unique in that it corresponds to the
base period of the squared sine function.  Other fractions would not have
the correct periodicity of the Hamiltonian.
The generalized overlaps for particle-number projection are then
\begin{subequations}
\begin{eqnarray}
  {\cal I}(\eta)
  &=&
  \exp({-2\langle\tilde{N}^2\rangle\sin^2\left(\frac{\eta}{2}\right)})
  \quad,
\\
  {\cal H}(\eta)
  &=&
  {\cal I}(\eta)\left[
    {\cal H}_0
    -
    \frac{1}{2}\sin^2\left(\frac{\eta}{2}\right){\cal H}_2
  \right]
  \quad,
\\
  {\cal H}_0
  &=&
  \langle\Phi_0|\hat{H}|\Phi_0\rangle
  =
  E_{\rm BCS}
  \quad,
\\
  {\cal H}_2
  &=&
  \langle\{\tilde{N},\{\hat{H}-\langle\hat{H}\rangle,\tilde{N}\}\}\rangle
  \quad.
\label{eq:H2pnp}
\end{eqnarray}
\end{subequations}
Note that the width $\lambda=2\langle\tilde{N}^2\rangle$ and the
coefficients ${\cal H}_i$ of the Hamiltonian overlap are still defined
as in standards GOA, see eq. (\ref{eq:GOA}). 
What changes is
the way how these overlaps are extrapolated. 
It is obvious that the conventional GOA is recovered in case of
steeply decaying norm overlap, i.e. for $\lambda\longrightarrow\infty$.

The projected energy (\ref{eq:Eproj}) can then be expressed in rather
compact fashion as
\begin{subequations}
\label{eq:PNPtop}
\begin{eqnarray}
  E
  &=&
  E_{\rm BCS}
  -
  \delta E_{\rm PNP}
  \quad,
\\
  \delta E_{\rm PNP}
  &=&
  \frac{1}{4}\Lambda\left(\langle\tilde{N}^2\rangle\right){\cal H}_2
  \quad,
\\
  \Lambda\left(y\right)
  &=&
  \frac{\int_0^{2\pi}d\eta\,
        e^{-2y\sin^2\left(\frac{\eta}{2}\right)}
        2\sin^2\left(\frac{\eta}{2}\right)}
       {\int_0^{2\pi}d\eta\,e^{-2y\sin^2\left(\frac{\eta}{2}\right)}} 
  \quad.
\end{eqnarray}
\end{subequations}
The limiting case standard GOA is recovered by 
$$
  \Lambda
  \longrightarrow
       1/(2\langle\tilde{N}^2\rangle) 
        \quad\mbox{for}\quad \langle\tilde{N}^2\rangle\rightarrow \infty
  \quad.
$$
This corresponds to a HFB state deep in the pairing regime where
one gathers substantial particle number fluctuations.
The opposite limit is
$$
  \Lambda
  \longrightarrow
       1 \quad\mbox{for}\quad \langle\tilde{N}^2\rangle\rightarrow 0
  \quad.
$$
It corresponds to the break down of pairing towards a pure HF
state. Standard GOA fails here. It is obvious that only topGOA can
cope properly with that pairing transition.


As in case of angular momentum projection, there is the choice between
variation-before-projection and variation-after-projection, see
section \ref{sec:VAP}. And again variation-after-projection is the
preferred method. Variation means here in general variation with
respect to the single particle wave functions in the HFB state and its
occupation amplitudes $u$ and $v$. The wave functions are fixed in the
model which we use later on and only the variation of $u$ and $v$
remains to be done.

\subsection{RPA correlations}

The correlation energy in RPA is computed with
eq. (\ref{eq:RPAcorr}). The mode corresponding to
particle-number phase is given by the path (\ref{eq:PNPpath}). It is
found as zero-energy mode in the RPA spectrum because of
$[\hat{H},\hat{N}]=0$. Thus one knows already the combination
$\hat{N}\equiv\hat{Q}=\left(\hat{C}^++\hat{C}\right)/\sqrt{2}$.  The
conjugate combination (\ref{eq:PQdef}) has to be determined by linear
response $[\hat{H},\hat{P}_N]\propto\hat{N}$. Once having the pair
$(\hat{N},\hat{P}_N)$, one can easily compute the correlation energy
(\ref{eq:RPAcorr}).

\subsection{A simple model as test case}

\subsubsection{The model}

For further testing of the approximate scheme, we need a schematic
model. It should have a gap in the single-particle spectrum to model
the interplay between this gap and the pairing strength.  Thus we take
a two-shell model with lower band $s=-1$ and upper band $s=+1$.  Each
band is $N$-fold degenerated as $m=-(N-1)/2,\cdots,+(N-1)/2$. The
states $\pm m$ are considered as the pairing conjugate partners. This
yields the generalized Lipkin-Meshkov-Glick model introduced in
Ref. \cite{feldman}. It is simply a two-level model with seniority
pairing. The model Hamiltonian reads
\begin{eqnarray}
  \hat{H}
  &=&
  \epsilon\sum_{sm}s\,\alpha_{sm}^+\alpha_{sm}^{\mbox{}}
\nonumber\\
  &&
  -\chi\frac{2\epsilon}{N}
  \left(\sum_{s,m\!>\!0}\alpha_{sm}^+\alpha_{s-m}^+\right)
  \left(\sum_{s,m\!>\!0}\alpha_{s-m}^{\mbox{}}\alpha_{sm}^{\mbox{}}\right)
  \quad.
\label{eq:LMG2_ham}
\end{eqnarray}
We associate the following
single-particle energies and occupation amplitudes
\begin{equation}
 \begin{array}{ll}
  \varepsilon_1 = \varepsilon
  &
  v_1 = u_{-1} = u = \sqrt{1-v^2}
  \\
  \varepsilon_{-1} = -\varepsilon
  &
  v_{-1} = u_1 = v
 \end{array}
\end{equation}
Note that the Fermi energy is $\varepsilon_F=0$ for symmetry reasons. 

The exact solution can be obtained by diagonalizing the Hamiltonian 
(\ref{eq:LMG2_ham}) using the quasi-spin formalism, for details see
\cite{HB00}.

\subsubsection{The energy in topGOA}

The model is sufficiently simple that everything can be worked out
analytically. The final result projected energy in topGOA becomes
\begin{subequations}
\begin{eqnarray}
  \frac{E_{\rm BCS}}{N\varepsilon}
  &=&
  -\left(\sqrt{1-(2uv)^2}+\frac{\chi}{2} (2uv)^2\right)
  \quad,
\\
  \delta E_{\rm PNP}
  &=&
  E_{\rm BCS} \left[1+(2uv)^2\Lambda\left(\frac{N}{2}(2uv)^2\right)\right]
  \quad.
\end{eqnarray}
\end{subequations}
This energy needs now to be compared with the BCS approximation
$E_{\rm BCS}$, the RPA energy, and the exact energy.

\subsubsection{The energy in RPA}

As shown in \cite{HB00}, there are two collective modes in this
model. For small values of $\chi$, the mean field approximation does
not support the BCS solution and only the trivial solution with zero 
pairing gap $\Delta=0$ appears. In this regime, the two RPA
frequencies are similar to each other, see Ref. \cite{HB00} for the
explicit expressions. 
At $\chi\sim N/(N-1)$, the system undergoes phase transition to 
the superfluid phase, and the number fluctuating BCS solution 
becomes the ground state in the mean field approximation. 
Consequently, one of the RPA frequencies becomes zero due to the
number conservation of the Hamiltonian (\ref{eq:LMG2_ham}). 
Applying eq.(\ref{eq:RPAcorr}) with the symmetry
mode yields the the RPA correlation energy
\begin{equation}
\Delta E_{\rm RPA}=-\frac{\epsilon\chi}{2}. 
\label{eq:PNP_RPA}
\end{equation}
The RPA frequency of the other mode is given by 
2$\Delta = 4\epsilon\chi(uv)$. This mode corresponds to the pairing vibration
whose contribution is omitted here because we study just the
projection part.

\begin{figure}[!tb]
\centerline{\epsfig{figure=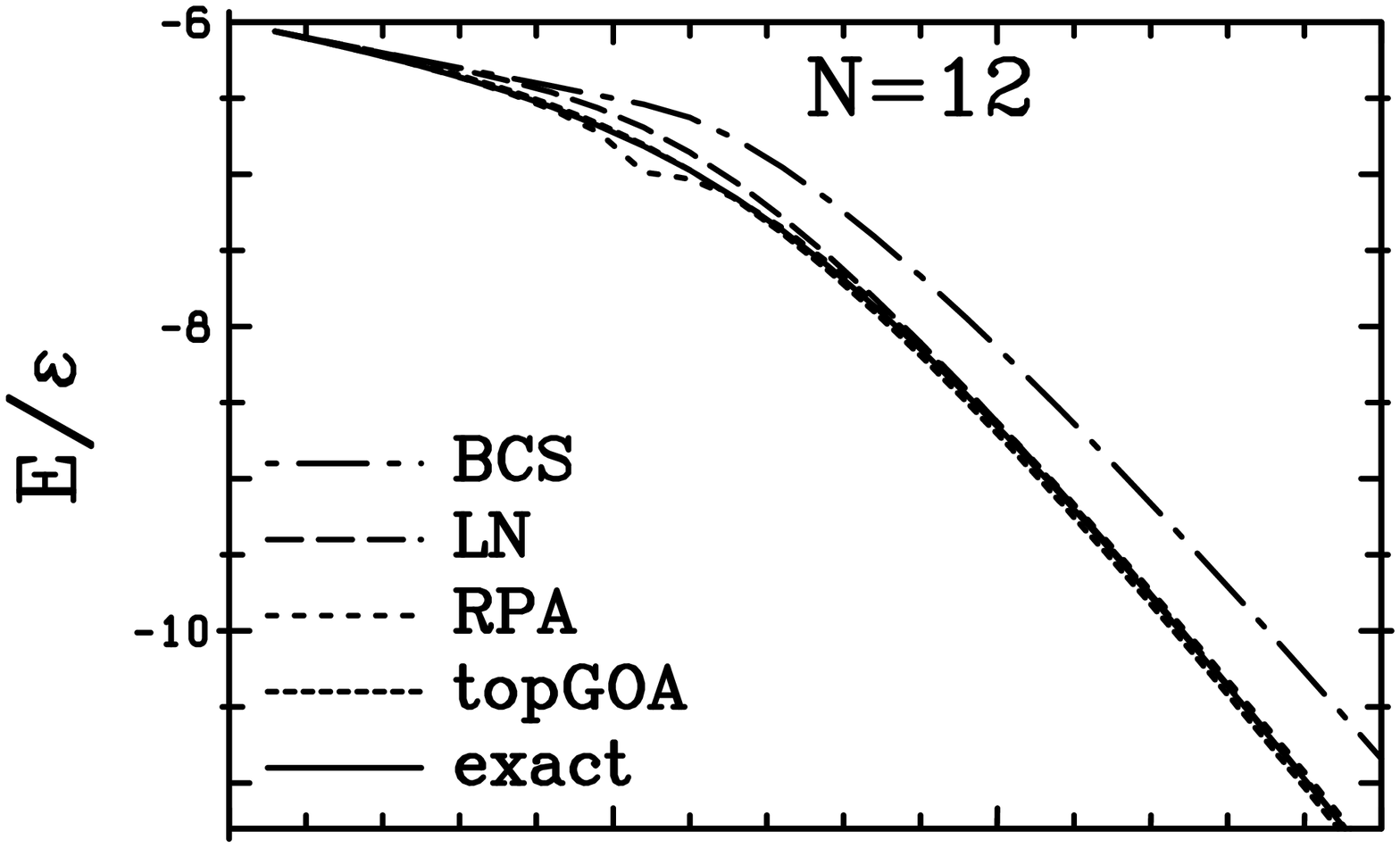,width=7.2cm,clip=}}
\centerline{\epsfig{figure=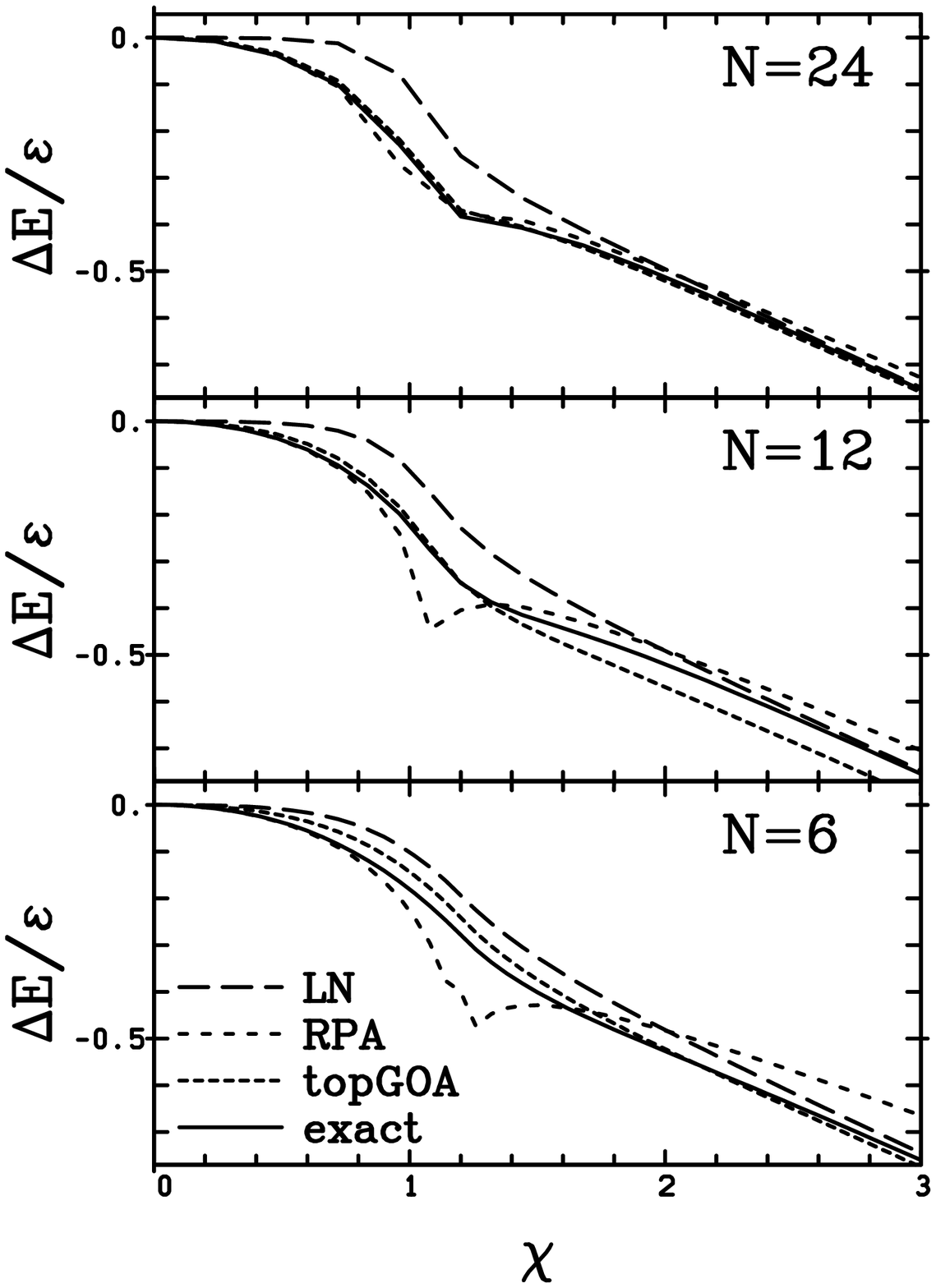,width=7.2cm,clip=}}
\caption{\label{fig:pnp_N12}\sl 
Upper panel: total energies at various levels of approximation
for a system with $N=12$ particles obeying the Hamiltonian
(\protect\ref{eq:LMG2_ham}).
Lower three panels:
The correlation energy $\Delta E =
E_{\rm BCS}$ for systems with
different $N$ as indicated.   }
\end{figure}

\subsubsection{A few words on the Lipkin-Nogami approach}
\label{sec:LN}

Full projection is often difficult, the more so if used in connection
with variation-after-projection. Thus one often employs approximate  schemes for
particle-number projection. 

A widely used approximation scheme for particle-number projection is
the Lipkin-Nogami approach, see e.g. \cite{Pra73a} and references
cited therein. It provides a good numerical approximation of the variation-after-projection
in situations where both the HFB equations predict a collapse of the
pairing correlations.  The prescription of Lipkin-Nogami amounts to
modify the energy by adding the second-order Kamlah correction
$\lambda_2(\hat{N}-\langle\hat{N}\rangle)^2$ where $\lambda_2$ is
computed from mixed variances of $\hat{N}$ and $\hat{H}$, see e.g.
for Skyrme-Hartree-Fock \cite{Rei96}.
The modification of the HFB equations associated with the Lipkin-Nogami
prescription is obtained by a restricted variation of where
$\lambda_2$ is not varied although its value is calculated from
self-consistent expectation values. For a thorough discussion of the
approximations involved see \cite{Flo97a, DN93}.
Note that the Kamlah expansion, and therefore the Lipkin-Nogami approach, uses
the similar expansion as the naive GOA and does not take into account
the topology of the gauge angle $\eta$.

\subsection{Results and discussion}
\label{sec:resPNP}

The upper part of figure \ref{fig:pnp_N12} shows the total energy in
the two-level model with seniority pairing for $N$=12 particles.  Various
approximations are considered. The BCS is the uncorrelated result.  It
decreases with constant slope up to 
$\chi\approx 1.1$ 
which
is the transition point from pure HF (for smaller $\chi$) to a truly
pairing HFB state (for larger $\chi$).  
The exact energy is the goal. In
addition to RPA and topGOA, we show also the results from the Lipkin-Nogami
scheme (see section \ref{sec:LN}). It is obvious from the figure
that all corrections improve the BCS energy towards the exact
result. The RPA correction works fine except for the region around the
critical point.  That is understandable because the critical point is
distinguished by large fluctuations and RPA is designed to be a theory
for small amplitude. The Lipkin-Nogami result has a smoother trend
than RPA and correct the energy in the wanted direction. However the
correction is incomplete, particularly at small coupling $\chi$ \cite{HB00}.
Last but not least, the topGOA provides a very good approximation
throughout all coupling strengths. It is clearly superior to the
competing projection approach, the Lipkin-Nogami scheme, and it is
more robust than RPA around the transition point.

A more detailed comparison of the various approaches is shown in the
lower panels of figure \ref{fig:pnp_N12}.  It displays the correlation
energies which point out the differences more clearly. First of all,
the correlation energy stays about independent of system size while
the total energy grows $\propto N$.  This means that the relative
importance of correlations shrinks as $1/N$. This corroborates the
known effect that mean field models, here represented by BCS, become
exact in the limit $N\longrightarrow\infty$.  The Lipkin-Nogami scheme
maintains its feature to produce a ``half-way'' correction. It is a
little bit surprising that the mismatch becomes even more pronounced with
increasing system size. The RPA, on the other hand, clearly improves
for larger systems. That is not surprising because mean field theories
are restored in the large $N$ limit, and RPA is a theory of vibrating
mean fields. Finally, the topGOA provides a reliable and robust
approximation to the exact correlation energy at all system sizes and
coupling strengths. There are regions where it is near to
perfect. There are regions where one obtains visible deviations of a
few percent. But the trends are always smooth and the average
performance is excellent.

There are two more  particularly appealing aspect of particle-number
projection with topGOA:
\begin{enumerate}
 \item The project energy (\ref{eq:PNPtop}) is a closed expression in
  terms of expectation values of $\hat{H}$ in combination with
  $\hat{N}$ and of the occupation amplitudes $u$ and $v$. One can
  easily use that as starting point for ``variation after projection''.
  Variation with respect to the single particle wave
  functions yields the appropriate correction terms to the mean field
  equations. These terms can easily be incorporated in existing codes.
  \item The full GCM is not applicable in connection with nuclear
  density functionals, as e.g. the Skyrme-Hartree-Fock
  energy. The energy-density functional is given for an expectation
  value with one mean field state. The extension to overlaps with
  different states at $q$ and $q'$ is ambiguous. But an extension of
  the functional is still feasible in the immediate vicinity of a mean
  field state. Thus the second order expression 
  ${\cal H}_2$ in (\ref{eq:H2pnp})   can still be derived
  within the safe grounds of density-functional theory. The topGOA
  thus provides a means to compute particle number projection safely
  for  Skyrme-Hartree-Fock.
\end{enumerate}

\section{Conclusions}

We have investigated the efficient computation of ground state
correlations for low-energy modes and projection. Starting point is
the generator coordinate method (GCM). It is considered in the
Gaussian overlap approximation (GOA) which reduces the formal and
numerical expense dramatically because it involves only expectation
values and second order variations therefrom. We have shown that GOA
runs into troubles in cases of weak coupling (thus broad overlaps) 
for coordinates with non-trivial topology. A slight modification of
the scheme allows to tune a topologically correct GOA (topGOA). 
We have demonstrated and tested topGOA for two typical cases of
collective coordinates, rotation-vibration and particle-number
projection. To that end, we employed exactly solvable models in the
spirit of the Lipkin-Meskov-Glick model.

The straightforward cases are mere projection (test cases: angular
momentum and particle number). It was found that topGOA provides an
excellent approximation to full projection. Performing variation after
projection (variation-after-projection) allows to incorporate already a great deal of
correlations into the projected states. The topGOA is particularly
well suited for variation-after-projection because the projected energy is expressed in
simple and compact expressions on which one can perform variation with
moderate expense, far simpler than for exact projection (where
non-orthogonal overlaps complicate matters). In particular for
particle number projection, topGOA thus offers a simple and in all
regimes reliable scheme which allows a thoroughly variational
formulation. It is superior to the Lipkin-Nogami scheme in that
respect.

Mere angular momentum projection with variation-after-projection was shown to grab a large
portion of the correlation energy. Yet it is incomplete without the
vibrational part. We have tested topGOA for the coupled
rotations-vibrations and it performs well in all regimes, near
sphericity, at the transition point, and for well deformed nuclei.  As
one could expect for a Gaussian limit, the performance improves with
system size. The reverse is also true: small systems are more critical
and a two-particle system is off limits. 

The topGOA for vibrations involves, in principle, large amplitude
motion. This can become inconvenient in practice because a whole
collective deformation path has to be mapped. The better defined
deformation of the variation-after-projection ground state allowed a small amplitude
expansion of topGOA. The result is a scheme which can live with second
order expression around the projected ground state. We consider it as
a topological generalization of the random-phase approximation (RPA)
which also deals with second order expressions throughout and call
this new scheme topological RPA (topRPA). We find that topRPA provides
a good approximation to the results of topGOA and thus to the exact
correlation energy for rotations-vibrations. 

Altogether, we have developed with the help of topologically corrected 
Gaussian overlaps a palette of useful approximations for computing
very efficiently collective correlations on top of nuclear mean-field
calculations. The next step is to implement that into practical
calculations. Work in that direction is in progress.

\section*{Acknowledgment}
This work was supported in part by
Bun\-des\-ministerium f\"ur Bildung und Forschung (BMBF), Project
No.\ 06 ER 808, by the Department of Energy under Grant
DE-FG06-90ER40561, and by the Grant-in-Aid for Scientific Research, 
Contract No. 12047203, from the Japanese Ministry of
Education, Culture, Sports, Science, and Technology.

\end{document}